# THE EAS BURSTS - HIGH ENERGY EXTENSIVE AIR SHOWERS CORRELATED IN TIME

T.T.Barnaveli, T.T.Barnaveli (jr), N.A.Eristavi, I.V.Khaldeeva
Andronikashvili Institute of Physics, Tamarashvili 6, Tbilisi, 0177, Georgia
A.P.Chubenko, N.M.Nesterova
Lebedev Physical Institute, Leninsky av., 53, Moscow, Russia

E-mail: tengiz.barnaveli@gmail.com

## ABSTRACT

Bursts of high energy EAS intensity (the “series” of EAS), following each other in short intervals of time were observed by means of Tien-Shan high mountain installation. For the lower boundary of EAS, uniting in one series, the size $N_e = 10^6$ (primary energies of the order of $4 \cdot 10^{15}$ eV) was taken. The condition of amalgamation into one series was the presence of at least two EAS of $N_e > 10^7$. The number of EAS in a series is from 4 to 9 events, with the mean time interval between them 1 – 5 minutes. Five such series were found in the material treated (approximately 250 days of pure time of the installation run). For each EAS of each series, all the basic parameters are given: observation date and time, age parameter S, galactic coordinates, coordinates of EAS axes relative to the installation center, energy release in the calorimeter and its distance from EAS axes, fitting the Nishimura-Kamata-Greisen (NKG) function. It is essential, that the frequency of the appearing of such series is much higher, than the probability of the occasional fluctuations in the succession of independent EAS. This probably is in favor of the commonality of their origin.

## 1. INTRODUCTION

In this article some results, obtained in the frames of our program of high and super high cosmic ray sources investigation are quoted. If the primary particles are gamma quanta, then the EAS axes orientation retains the direction to the source. If the primary particles are not gamma quanta, identification of the source by means of the registered EAS axes orientation distribution (even of very high energies) seems to be an uncertain task. The scattering of the primary particles on their way from the source to the Earth is too high, mainly due to deviation in the galactic magnetic fields. So we decided to base our analysis on the correlation of the events in time. That means that the “series” of high energy events, following each other in short intervals of time (<~ 30 min.), were searched for and selected. For the lower boundary of EAS in series the size $N_e = 10^6$ (primary energies of the order of $4 \cdot 10^{15}$ eV) was taken. The condition of amalgamation into one series was the presence of at least two EAS of $N_e > 10^7$. There were 5 such series found in the handled material (about 250 days pure time of run of the installation).

The data were obtained by means of the Tien-Shan high mountain installation in the period before 1980, and were retreated a new in the frames of the new tasks and approach. More than 300 000 events were handled in total.

The installation contained the scintillation detectors aimed at EAS extraction and evaluation of their size – in the center of the installation, and numerous scintillation transducers located symmetrically at the distances 15 m and 20 m from the center, as well the transducers at distance 73 m from the center, and the “carpet” of 64 scintillation transducers aimed at study of the central part of the EAS in details. The hadron component was studied by means of the large ionization calorimeter, containing from 16 to 20 layers (in this experiment the number of layers was 16 ) of ionization chambers. The area of each layer is 36 $m^2$. There were muon and other transducers as well. The detailed description of the installation can be found in [1] for example, while the data bank is described in [2]. The triggering conditions and the general principles of data treatment are described in [3] and in the works quoted there.

## 2. EXPERIMENTAL RESULTS.

The data concerning the series found are given below, including the basic graphical material. For each of the series two graphs are given. On one of the graphs the position of the series on the dependence $N_e(t)$ (built for the run under consideration) is represented. On the second graph is shown the distribution of the galactic coordinates of EAS axes directions for the EAS composing the given series. The graphs are followed by Tables, where the events included in the series are highlighted in bold. For each EAS of each series all the basic parameters are given: observation date and time (in hours and decimal parts of hours, according the 24 hour system, the minimal registered interval of time is one minute, LT=GMT+6 hours), the EAS size $N_e$ and corresponding primary energy $E_0$ (eV), age parameter S, galactic coordinates L and B of each EAS, energy release in the calorimeter (the specific - per one chamber - release EK of hadron component energy, arbitrary units), the coordinates X and Y of EAS axes relatively to the center of the installation and the distance R of the energy release center of weight from EAS axis, fit level FF to NKG function (square of relative deviation from NKG distribution over all transducers of the installation). The age parameter S is defined by means of all transducers of the installation as a whole, i.e. S is regarded as a formal parameter, and in the fitting process could obtain the values in the limits 0.01 – 1.99. For each series the galactic coordinates $\langle L\rangle$ and $\langle B\rangle$ of the centre of weight was estimated – using the EAS primary energy $E_0$ for the weight. The numbers assigned to the series correspond to the numbers of the figures where they are represented. The density of EAS distribution on graphs is defined by triggering conditions and registration threshold, which could differ in different runs and sometimes could be changed in the frames of one run, dependent on the task set. However the influence of these input conditions is limited by the region $N_e <\sim 5 \cdot 10^5$, which is by about 1 – 4 orders lower than the values characteristic for the series observed. The gaps in observation, seen on some of the graphs, correspond to pauses assigned for service works.

The series are represented in order of the calendar dates and time of registration.

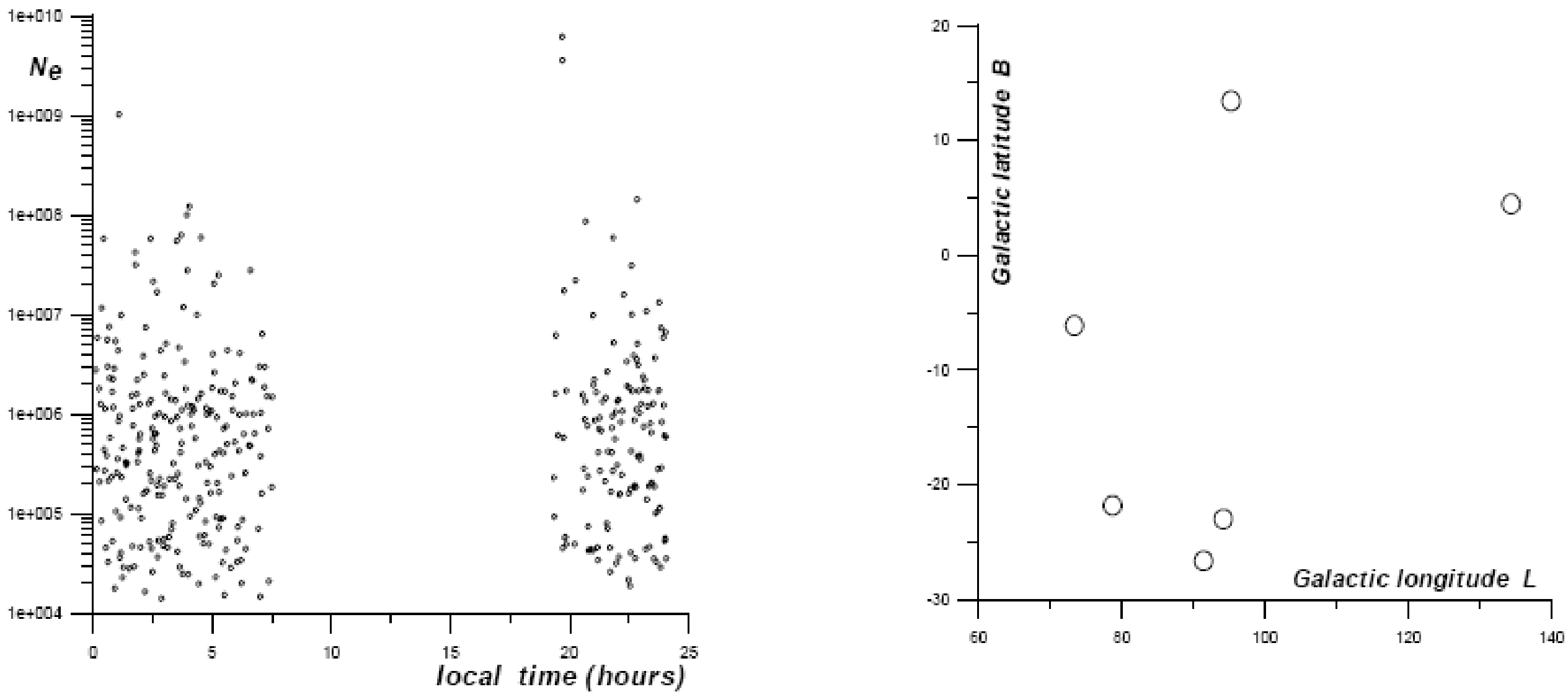

Fig.1. Series № 1 (Run № 2342), 6 EAS, 21.11.1973, <B> = -2.14°, <L>=99.74°, direction of maximal scattering 34.5° to the Galaxy equator plane. Probability of the occasional observation is $< 10^{-11}$.

Table of data for Series № 1 (Run № 2342)

| LOC. TIME | $N_E$ | $E_0$ | B | L | S | EK | FF | X | Y | R |
|---|---|---|---|---|---|---|---|---|---|---|
| **19.42** | **$3.04 \cdot 10^6$** | **$6.56 \cdot 10^{15}$** | **-23.0** | **94.18** | **1.518** | **$1.7 \cdot 10^{-5}$** | **$3.6 \cdot 10^{-2}$** | **-95.5** | **51.8** | **106.9** |
| 19.42 | $1.15 \cdot 10^5$ | $3.80 \cdot 10^{14}$ | -37.19 | 113.87 | 0.855 | 0.48 | $3.9 \cdot 10^{-2}$ | 15.6 | 3.7 | 12.0 |
| **19.43** | **$6.12 \cdot 10^6$** | **$1.20 \cdot 10^{16}$** | **4.45** | **134.36** | **1.952** | **$1.4 \cdot 10^{-5}$** | **$2.95 \cdot 10^{-2}$** | **-18.1** | **51.9** | **39.7** |
| 19.53 | $6.06 \cdot 10^5$ | $1.60 \cdot 10^{15}$ | -8.39 | 98.88 | 1.68 | $1.8 \cdot 10^{-5}$ | $4.69 \cdot 10^{-2}$ | -16.3 | 27.1 | 31.1 |
| **19.70** | **$6.09 \cdot 10^9$** | **$4.88 \cdot 10^{18}$** | **13.42** | **95.26** | **0.171** | **0.52** | **$7.29 \cdot 10^{-2}$** | **-2.8** | **-10.8** | **12.3** |
| **19.72** | **$3.55 \cdot 10^9$** | **$3.05 \cdot 10^{18}$** | **-26.66** | **91.43** | **0.202** | **0.495** | **$7.32 \cdot 10^{-2}$** | **3.2** | **-15.6** | **14.2** |
| 19.73 | $4.48 \cdot 10^4$ | $1.67 \cdot 10^{14}$ | -10.5 | 86.32 | 0.574 | 0.59 | 0.697 | 4.9 | 8.2 | 10.2 |
| 19.77 | $5.74 \cdot 10^5$ | $1.54 \cdot 10^{15}$ | 10.28 | 100.78 | 0.671 | $1.2 \cdot 10^{-2}$ | $5.38 \cdot 10^{-2}$ | -29.9 | 45.0 | 49.7 |
| **19.78** | **$3.44 \cdot 10^7$** | **$5.40 \cdot 10^{16}$** | **-21.83** | **78.72** | **0.438** | **$3.0 \cdot 10^{-3}$** | **$5.05 \cdot 10^{-3}$** | **-148.5** | **5.1** | **145.3** |
| 19.78 | $5.19 \cdot 10^4$ | $1.89 \cdot 10^{14}$ | -3.7 | 111.45 | 0.975 | 0.58 | $3.69 \cdot 10^{-2}$ | -3.3 | -2.0 | 3.2 |
| 19.82 | $6.65 \cdot 10^4$ | $2.04 \cdot 10^{14}$ | 6.54 | 129.12 | 1.199 | $4.9 \cdot 10^{-3}$ | $7.2 \cdot 10^{-2}$ | 1.2 | 3.8 | 3.2 |
| 19.83 | $5.76 \cdot 10^4$ | $2.08 \cdot 10^{14}$ | 3.22 | 119.87 | 1.23 | 0.79 | $4.66 \cdot 10^{-2}$ | -4.1 | 1.9 | 1.7 |
| **19.87** | **$1.09 \cdot 10^6$** | **$3.95 \cdot 10^{15}$** | **-6.15** | **73.38** | **0.91** | **$5.6 \cdot 10^{-2}$** | **$3.87 \cdot 10^{-2}$** | **-15.1** | **-13.2** | **16.1** |

As an example, for the series № 1 the picture of the energy release in the ionization calorimeter is represented (the other series are not shown for reasons of article volume economy, but the represented picture qualitatively does not differ from other series). For each event, both projections of the calorimeter are shown. The mutual position of the layers is kept up. Each event of the series is marked by the time of its registration, in accordance with the left column of the Table. All events with the nonzero energy release are shown. The energy release EK of the order of $\sim 10^{-5}$ is taken as zero.

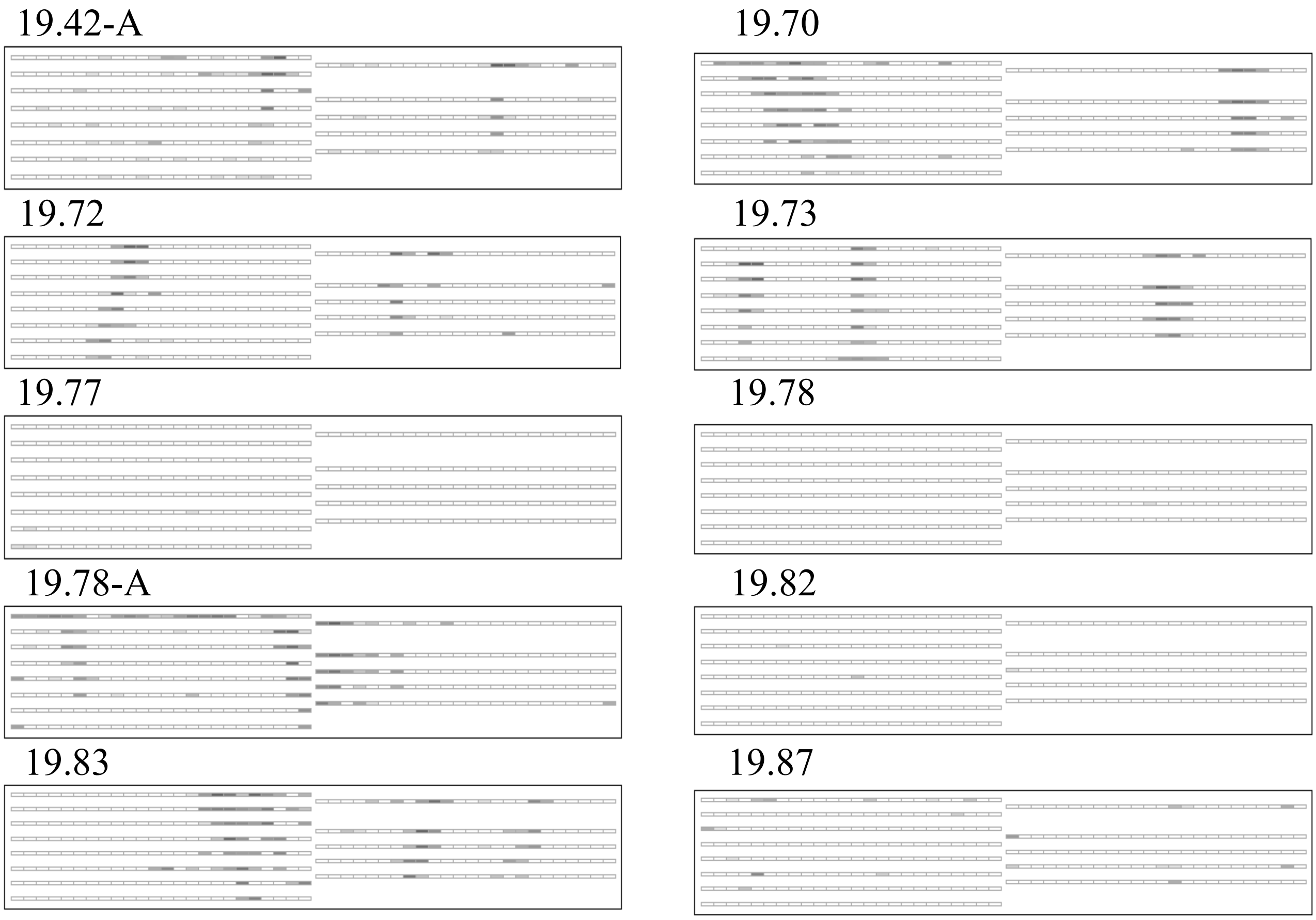

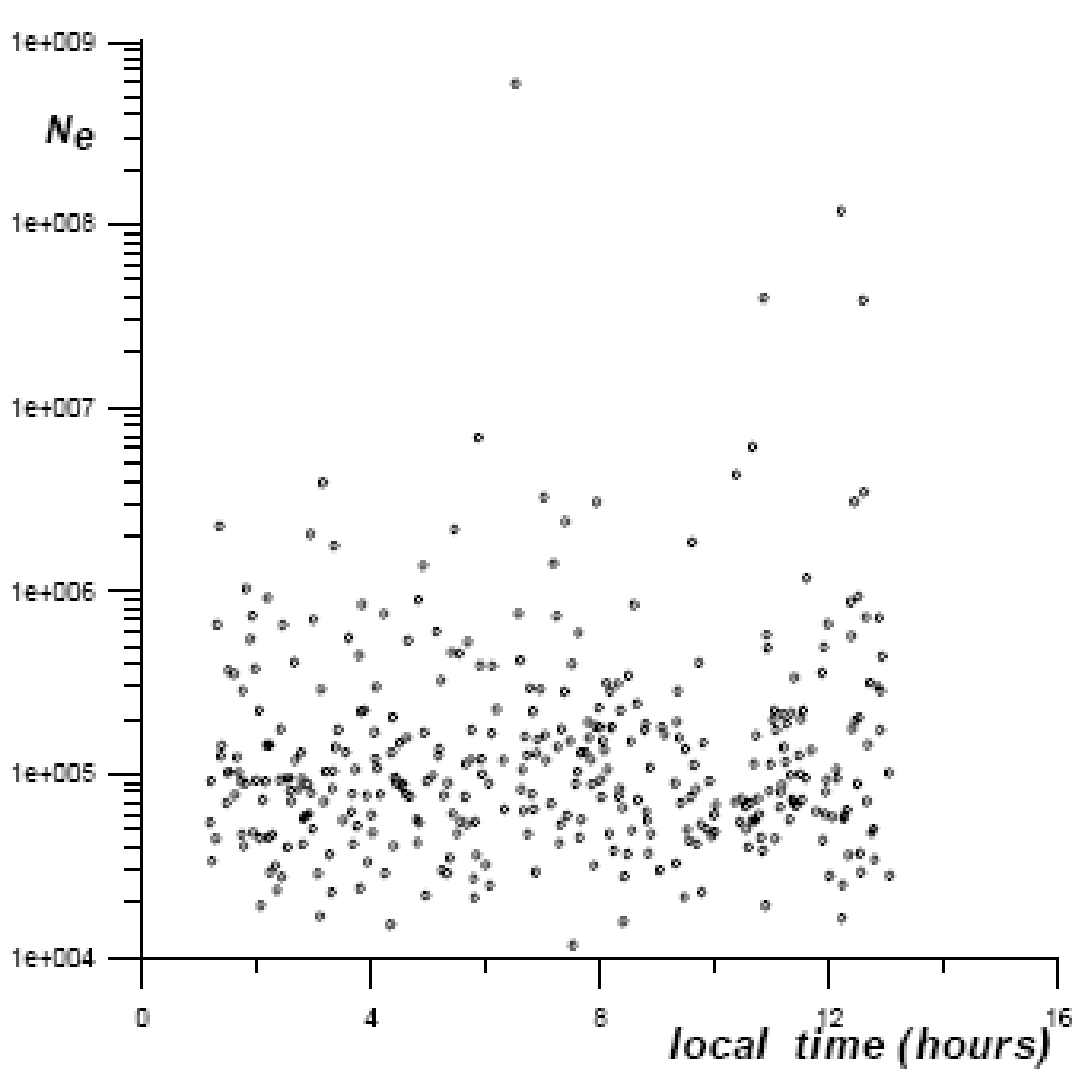

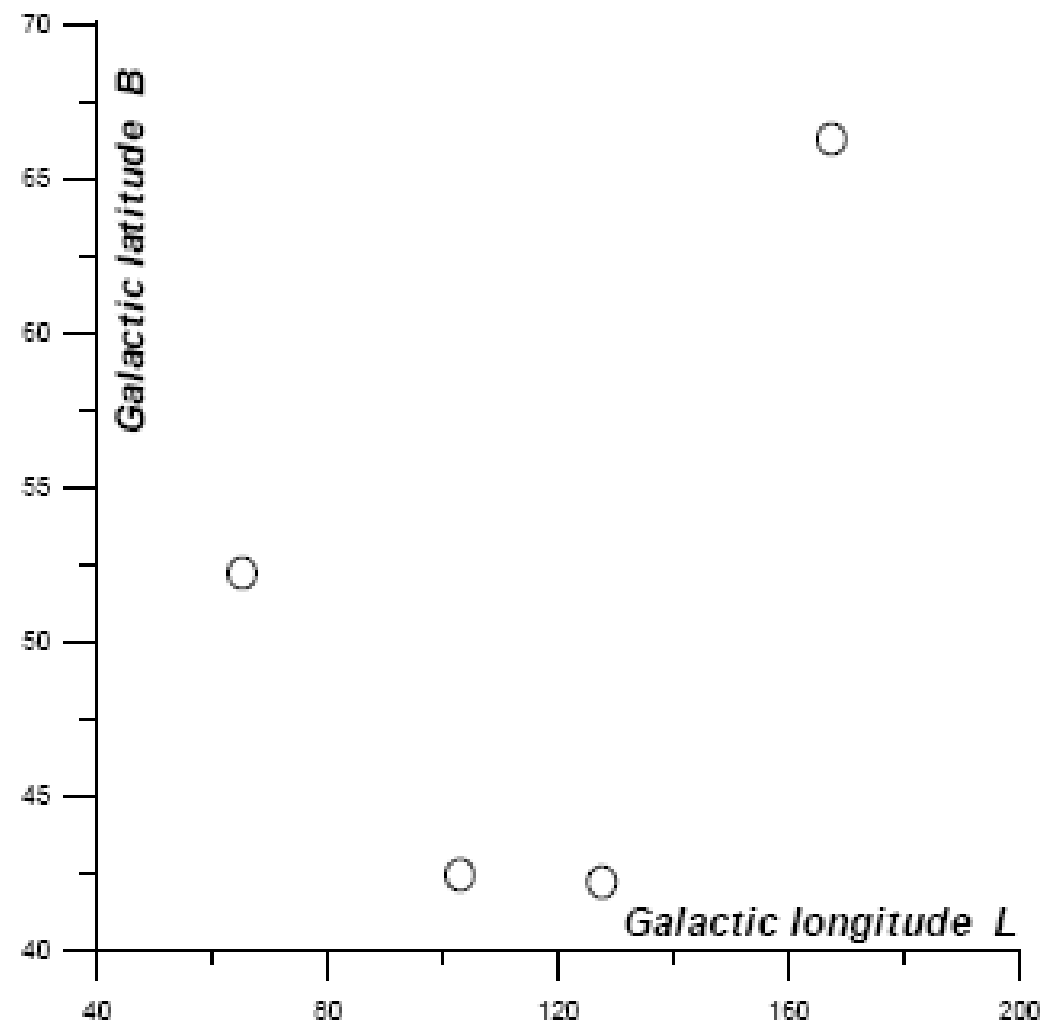

Fig.2. Series № 2 (Run № 2746), 4 EAS, 12.10.1974, <B>=58.57°, <L>=151.07°, direction of maximal scattering 10.5° to the Galaxy equator plane. Probability of the occasional observation is < $2.5 \cdot 10^{-2}$.

Table of data for series № 2 (Run № 2746)

| LOC. TIME | $N_E$ | $E_0$ | B | L | S | EK | FF | X | Y | R |
|---|---|---|---|---|---|---|---|---|---|---|
| **12.22** | **$1.19 \bullet 10^{8}$** | **$1.59 \bullet 10^{17}$** | **66.25** | **167.39** | **0.43** | **1.33** | **$3.55 \bullet 10^{-2}$** | **-11.7** | **-27.6** | **27.1** |
| 12.23 | $1.64 \bullet 10^{4}$ | $6.95 \bullet 10^{13}$ | 53.23 | 81.33 | 0.70 | 0.26 | $6.03 \bullet 10^{-2}$ | 2.6 | 7.9 | 2.4 |
| 12.25 | $2.48 \bullet 10^{4}$ | $9.99 \bullet 10^{13}$ | 84.57 | 69.59 | 0.79 | 0.44 | $5.25 \bullet 10^{-2}$ | 6.1 | 5.7 | 4.0 |
| 12.26 | $5.74 \bullet 10^{4}$ | $2.07 \bullet 10^{14}$ | 77.74 | 45.13 | 0.76 | $1.66 \bullet 10^{-2}$ | $3.40 \bullet 10^{-2}$ | 3.8 | 6.7 | 7.1 |
| 12.28 | $5.98 \bullet 10^{4}$ | $2.15 \bullet 10^{14}$ | 77.71 | 126.11 | 0.95 | $3.23 \bullet 10^{-3}$ | $3.18 \bullet 10^{-2}$ | 4.5 | 3.0 | 7.0 |
| 12.32 | $6.37 \bullet 10^{4}$ | $2.27 \bullet 10^{14}$ | 45.49 | 121.61 | 1.01 | 0.52 | $5.11 \bullet 10^{-2}$ | 5.7 | -1.5 | 2.8 |
| 12.35 | $3.02 \bullet 10^{4}$ | $1.18 \bullet 10^{14}$ | 33.65 | 110.03 | 0.57 | 1.39 | $4.29 \bullet 10^{-2}$ | 3.3 | -5.9 | 7.2 |
| 12.35 | $4.21 \bullet 10^{4}$ | $1.58 \bullet 10^{14}$ | 62.44 | 114.01 | 0.60 | $2.94 \bullet 10^{-2}$ | $4.21 \bullet 10^{-2}$ | 10.3 | -3.3 | 5.3 |
| 12.38 | $8.73 \bullet 10^{5}$ | $2.21 \bullet 10^{15}$ | 83.75 | 129.56 | 1.95 | $3.19 \bullet 10^{-3}$ | $5.76 \bullet 10^{-2}$ | -5.3 | 7.9 | 9.7 |
| 12.39 | $9.54 \bullet 10^{5}$ | $2.39 \bullet 10^{15}$ | 69.90 | 126.07 | 0.80 | 0.15 | $1.97 \bullet 10^{-2}$ | 9.4 | -8.6 | 9.9 |
| 12.39 | $1.79 \bullet 10^{5}$ | $5.58 \bullet 10^{14}$ | 80.27 | 140.05 | 1.09 | $8.01 \bullet 10^{-2}$ | $2.99 \bullet 10^{-2}$ | -2.8 | -5.1 | 4.3 |
| 12.42 | $1.77 \bullet 10^{5}$ | $5.52 \bullet 10^{14}$ | 46.65 | 68.60 | 1.30 | $3.46 \bullet 10^{-2}$ | $3.48 \bullet 10^{-2}$ | -5.4 | 12.9 | 11.2 |
| **12.45** | **$6.05 \bullet 10^{6}$** | **$1.19 \bullet 10^{16}$** | **42.41** | **102.95** | **0.75** | **0.128** | **$1.76 \bullet 10^{-2}$** | **-21.9** | **3.9** | **18.1** |
| 12.45 | $4.83 \bullet 10^{4}$ | $2.72 \bullet 10^{14}$ | 68.04 | 200.35 | 1.56 | 0.24 | $8.29 \bullet 10^{-2}$ | 1.5 | -3.1 | 2.4 |
| 12.46 | $1.96 \bullet 10^{5}$ | $6.03 \bullet 10^{14}$ | 63.44 | 178.32 | 0.71 | 0.19 | $2.26 \bullet 10^{-2}$ | -4.5 | -7.1 | 4.5 |
| 12.49 | $8,84 \bullet 10^{4}$ | $3.01 \bullet 10^{15}$ | 85.39 | 185.51 | 0.44 | 0.12 | $3.29 \bullet 10^{-2}$ | -1.7 | -10.5 | 8.8 |
| 12.52 | $9.33 \bullet 10^{5}$ | $2.34 \bullet 10^{14}$ | 81.02 | 98.73 | 1.93 | $1.75 \bullet 10^{-5}$ | $6.58 \bullet 10^{-2}$ | -2.2 | 5.0 | 5.6 |
| 12.53 | $2.23 \bullet 10^{5}$ | $6.74 \bullet 10^{14}$ | 50.44 | 34.07 | 1.17 | $7.83 \bullet 10^{-3}$ | $3.00 \bullet 10^{-2}$ | -9.7 | 3.6 | 10.2 |
| 12.53 | $1.83 \bullet 10^{5}$ | $5.69 \bullet 10^{14}$ | 43.23 | 153.17 | 0.91 | 0.328 | $2.46 \bullet 10^{-2}$ | 3.3 | 4.5 | 4.1 |
| 12,54 | $3.69 \bullet 10^{4}$ | $1.41 \bullet 10^{14}$ | 44.50 | 97.26 | 0.50 | $1.39 \bullet 10^{-2}$ | $4.87 \bullet 10^{-2}$ | 0.7 | -9.9 | 9.0 |
| 12.56 | $2.91 \bullet 10^{4}$ | $1.15 \bullet 10^{14}$ | 72.75 | 84.06 | 0.84 | 0.29 | $5.00 \bullet 10^{-2}$ | 1.1 | 2.9 | 2.3 |
| **12.60** | **$3.87 \bullet 10^{7}$** | **$5.99 \bullet 10^{16}$** | **42.17** | **127.57** | **0.129** | **$5.84 \bullet 10^{-2}$** | **$4.01 \bullet 10^{-2}$** | **-37.1** | **-16.3** | **35.8** |
| **12.62** | **$3.46 \bullet 10^{6}$** | **$7.32 \bullet 10^{15}$** | **52.17** | **65.21** | **0.82** | **2.51** | **$2.67 \bullet 10^{-2}$** | **3.5** | **-4.2** | **6.0** |

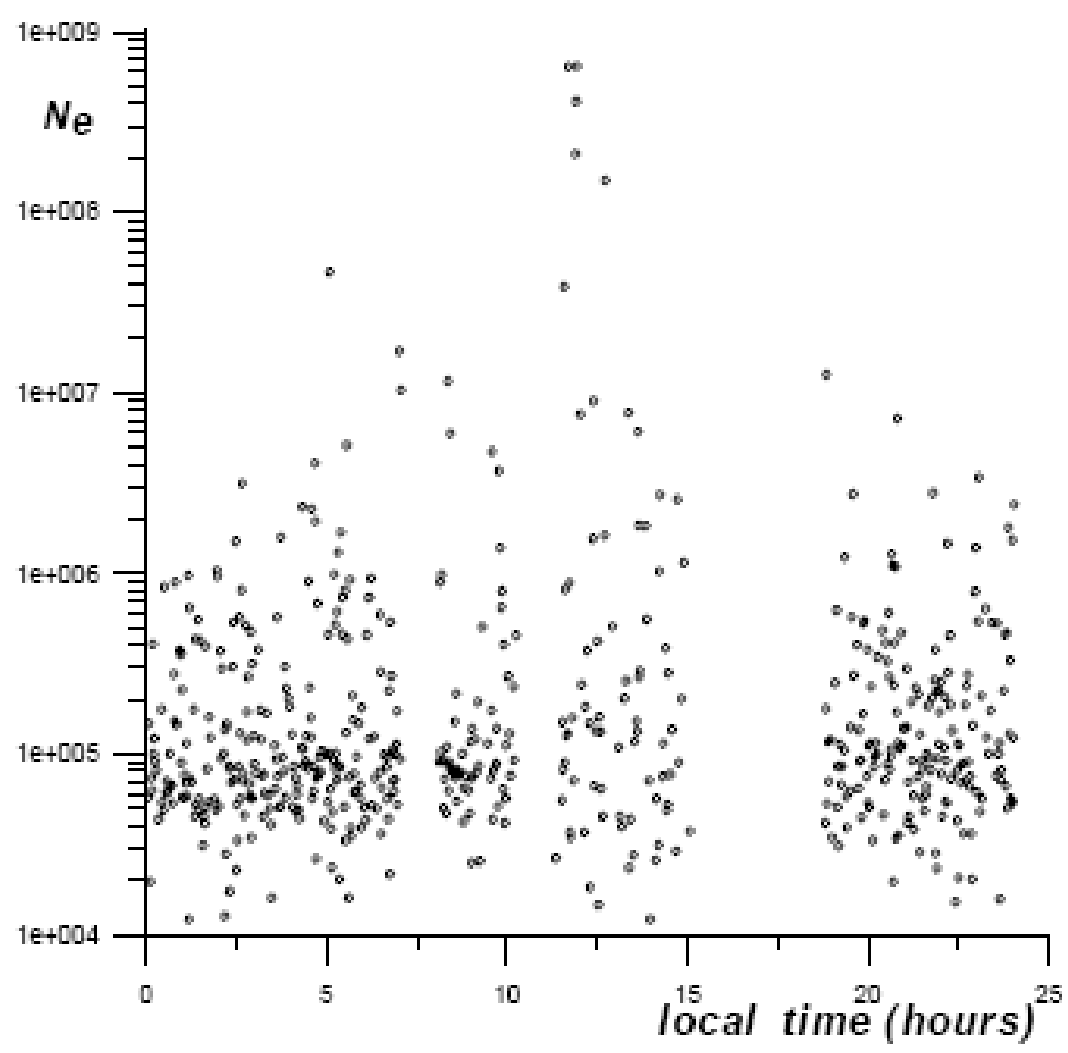

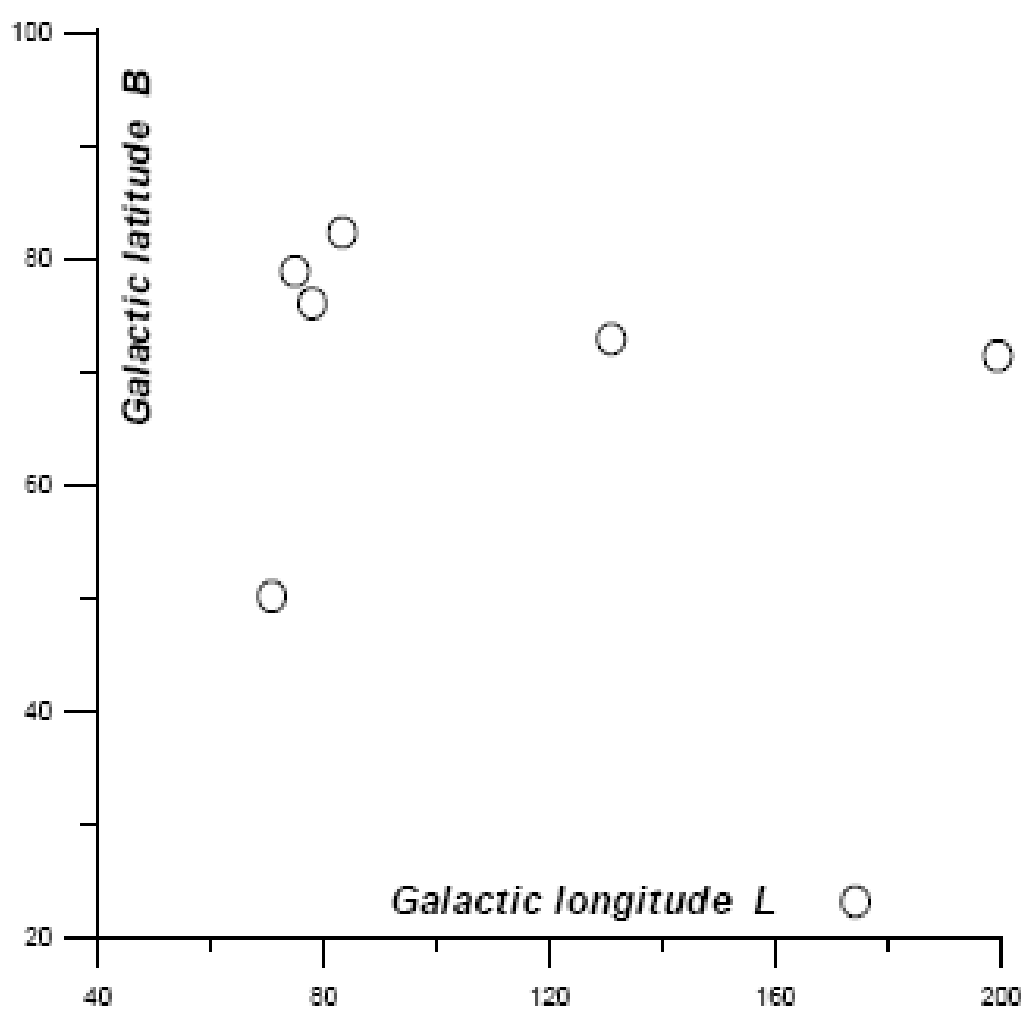

Fig. 3. Series № 3 (Run № 2751), 9 EAS, 15.10.1974, <B>=60.47°, <L>=132.45°, direction of maximal scattering 170.5° to the Galaxy equator plane. Probability of the occasional observation is < $2 \bullet 10^{-17}$.

Table of data for series № 3 (Run № 2751)

| LOC. TIME | $N_E$ | $E_0$ | B | L | S | EK | FF | X | Y | R |
|---|---|---|---|---|---|---|---|---|---|---|
| **11.59** | **$3.87 \cdot 10^{7}$** | **$4.4 \cdot 10^{16}$** | **58.25** | **87.07** | **0.113** | **0.386** | **$6.6 \cdot 10^{-2}$** | **12.5** | **1.1** | **13.4** |
| 11.62 | $8.94 \cdot 10^{4}$ | $3.05 \cdot 10^{14}$ | 32.05 | 122.81 | 0.131 | 0.304 | $8.2 \cdot 10^{-2}$ | -3.8 | 6.8 | 4.0 |
| 11.63 | $8.12 \cdot 10^{5}$ | $2.08 \cdot 10^{15}$ | 76.92 | 135.54 | 1.882 | $1.5 \cdot 10^{-5}$ | $4.7 \cdot 10^{-2}$ | -9.5 | 3.6 | 10.2 |
| 11.67 | $2.88 \cdot 10^{4}$ | $1.14 \cdot 10^{14}$ | 79.18 | 126.17 | 1.084 | 0.342 | $5.3 \cdot 10^{-2}$ | 1.5 | -2.2 | 3.7 |
| 11.67 | $3.64 \cdot 10^{4}$ | $1.39 \cdot 10^{14}$ | 53.22 | 40.72 | 0.997 | 1.071 | $4.0 \cdot 10^{-2}$ | 2.6 | 1.4 | 1.8 |
| 11.67 | $3.13 \cdot 10^{5}$ | $9.07 \cdot 10^{14}$ | 56.72 | 141.1 | 0.873 | 0.295 | $2.5 \cdot 10^{-2}$ | -4.8 | 3.2 | 5.1 |
| 11.69 | $1.29 \cdot 10^{5}$ | $4.21 \cdot 10^{14}$ | 69.04 | 163.52 | 0.817 | 0.253 | $2.6 \cdot 10^{-2}$ | -3.2 | -1.8 | 0.9 |
| **11.72** | **$6,39 \cdot 10^{8}$** | **$6.86 \cdot 10^{17}$** | **23.07** | **174.15** | **0.176** | **0.587** | **$3.6 \cdot 10^{-2}$** | **8.3** | **-36.9** | **36.0** |
| **11.75** | **$2.58 \cdot 10^{6}$** | **$5.68 \cdot 10^{15}$** | **50.07** | **70.85** | **0.732** | **0.155** | **$1.6 \cdot 10^{-2}$** | **-5.9** | **22.6** | **15.1** |
| 11.75 | $3.32 \cdot 10^{4}$ | $1.29 \cdot 10^{14}$ | 78.97 | 207.56 | 1.213 | 0.182 | $6.2 \cdot 10^{-2}$ | 4.3 | -2.0 | 4.5 |
| 11.75 | $2.78 \cdot 10^{4}$ | $1.10 \cdot 10^{14}$ | 76.22 | 103.14 | 0.869 | 0.445 | $5.9 \cdot 10^{-2}$ | -1.1 | 0.5 | 2.2 |
| 11.76 | $3.69 \cdot 10^{4}$ | $1.41 \cdot 10^{14}$ | 54.52 | 181.39 | 0.586 | 0.291 | $6.0 \cdot 10^{-2}$ | 7.2 | -6.4 | 7.6 |
| 11.78 | $3.44 \cdot 10^{4}$ | $1.33 \cdot 10^{14}$ | 55.58 | 134.05 | 0.808 | 0.283 | $5.3 \cdot 10^{-2}$ | 3.9 | -7.9 | 1.4 |
| 11.83 | $1.58 \cdot 10^{5}$ | $4.99 \cdot 10^{14}$ | 77.16 | 69.99 | 0.633 | 0.618 | $2.0 \cdot 10^{-2}$ | 4.5 | 6.5 | 5.2 |
| 11.86 | $7.12 \cdot 10^{4}$ | $2.50 \cdot 10^{14}$ | 81.02 | 103.39 | 0.512 | 0.248 | $3.4 \cdot 10^{-2}$ | -6.4 | 1.6 | 6.0 |
| **11.92** | **$1.68 \cdot 10^{8}$** | **$1.5 \cdot 10^{17}$** | **75.94** | **77.96** | **0.887** | **0.112** | **$7.49 \cdot 10^{-2}$** | **-1.3** | **0.9** | **1.5** |
| **11.92** | **$2.52 \cdot 10^{8}$** | **$2.15 \cdot 10^{17}$** | **78.81** | **74.93** | **0.420** | **0.214** | **$7.62 \cdot 10^{-2}$** | **1.0** | **3.7** | **3.2** |
| **11.93** | **$5.68 \cdot 10^{8}$** | **$4.25 \cdot 10^{17}$** | **82.22** | **83.33** | **0.087** | **$1.68 \cdot 10^{-5}$** | **$7.96 \cdot 10^{-2}$** | **1.1** | **-0.4** | **2.4** |
| **11.93** | **$2.52 \cdot 10^{8}$** | **$2.15 \cdot 10^{17}$** | **71.31** | **199.36** | **1.057** | **0.010** | **$7.44 \cdot 10^{-2}$** | **-0.2** | **1.5** | **4.6** |
| **11.95** | **$6.41 \cdot 10^{8}$** | **$4.7 \cdot 10^{17}$** | **72.83** | **130.91** | **1.159** | **$1.75 \cdot 10^{-5}$** | **$7.17 \cdot 10^{-2}$** | **-2.5** | **0.02** | **2.8** |
| **12.04** | **$7.56 \cdot 10^{6}$** | **$1.45 \cdot 10^{17}$** | **75.45** | **142.82** | **0.374** | **$1.76 \cdot 10^{-5}$** | **$3.45 \cdot 10^{-2}$** | **-66.6** | **36.6** | **75.4** |

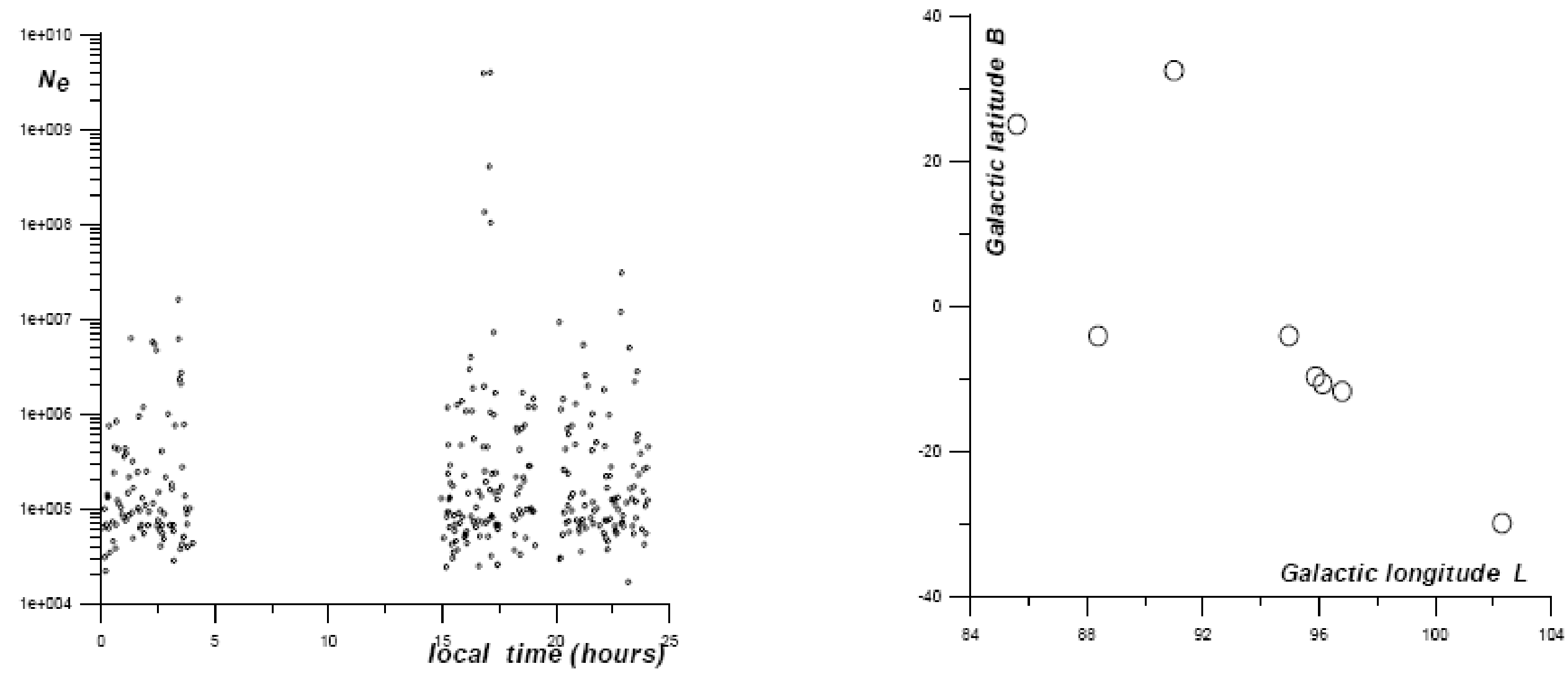

Fig.4. Series № 4 (Run № 2855), 8EAS, 25.12.1974, $\langle B \rangle = -6.05^{o}$, $\langle L \rangle = 79.69^{o}$, direction of maximal scattering $94.5^{o}$ to the Galaxy equator plane. Probability of the occasional observation is $< 10^{-19}$.

Table of data for series № 4 (Run № 2855)

| LOC. TIME | $N_E$ | $E_0$ | B | L | S | EK | FF | X | Y | R |
|---|---|---|---|---|---|---|---|---|---|---|
| **16.83** | **$3.86 \cdot 10^{9}$** | **$3.28 \cdot 10^{18}$** | **-4.15** | **94.97** | **0.103** | **2.45** | **$7.8 \cdot 10^{-2}$** | **1.5** | **1.6** | **3.0** |
| **16.85** | **$3.49 \cdot 10^{6}$** | **$7.39 \cdot 10^{15}$** | **32.42** | **91.01** | **0.058** | **0.046** | **$5.9 \cdot 10^{-2}$** | **5.0** | **-5.7** | **7.4** |
| 16.85 | $3.67 \cdot 10^{5}$ | $1.04 \cdot 10^{15}$ | -28.36 | 133.02 | 0.154 | 0.005 | $3.4 \cdot 10^{-2}$ | 11.7 | 8.4 | 9.8 |
| **16.87** | **$2.66 \cdot 10^{8}$** | **$3.20 \cdot 10^{17}$** | **-4.17** | **88.39** | **0.816** | **3.10** | **$7.6 \cdot 10^{-2}$** | **-1.2** | **-1.7** | **2.9** |
| 16.87 | $1.42 \cdot 10^{5}$ | $4.55 \cdot 10^{14}$ | -12.02 | 91.47 | 0.798 | 0.653 | $2.9 \cdot 10^{-2}$ | 5.6 | -6.9 | 9.4 |
| 16.88 | $2.47 \cdot 10^{5}$ | $7.37 \cdot 10^{14}$ | -24.07 | 89.23 | 0.650 | 0.210 | $2.0 \cdot 10^{-2}$ | 2.8 | -8.1 | 8.6 |
| 16.93 | $2.51 \cdot 10^{5}$ | $7.47 \cdot 10^{14}$ | 7.76 | 91.73 | 0.606 | 0.122 | $3.6 \cdot 10^{-2}$ | -13.3 | -5.6 | 11.6 |
| 16.93 | $1.29 \cdot 10^{5}$ | $4.18 \cdot 10^{14}$ | -25.01 | 52.59 | 0.320 | $1.3 \cdot 10^{-5}$ | $3.3 \cdot 10^{-2}$ | -12.5 | 6.7 | 13.1 |
| 16.95 | $6.18 \cdot 10^{5}$ | $2.21 \cdot 10^{14}$ | -8.92 | 136.11 | 0.889 | 0.074 | $3.5 \cdot 10^{-2}$ | -1.4 | -5.2 | 7.7 |
| 16.95 | $7.94 \cdot 10^{4}$ | $2.75 \cdot 10^{14}$ | 2.66 | 86.66 | 0.881 | 0.001 | $2.2 \cdot 10^{-2}$ | -2.0 | 6.7 | 5.1 |
| 16.99 | $2.14 \cdot 10^{4}$ | $6.51 \cdot 10^{14}$ | 8.68 | 75.27 | 1.24 | $1.6 \cdot 10^{-3}$ | $2.6 \cdot 10^{-2}$ | -10.3 | 0.01 | 6.1 |
| 16.99 | $6.72 \cdot 10^{5}$ | $1.76 \cdot 10^{15}$ | -2.27 | 90.28 | 1.91 | $9.5 \cdot 10^{-3}$ | $6.0 \cdot 10^{-2}$ | 4.4 | 13.2 | 15.7 |
| 17.03 | $5.04 \cdot 10^{5}$ | $1.85 \cdot 10^{14}$ | -13.97 | 101.84 | 0.571 | 0.232 | $4.5 \cdot 10^{-2}$ | -5.1 | -0.9 | 4.0 |
| 17.05 | $7.59 \cdot 10^{4}$ | $2.64 \cdot 10^{14}$ | -1.88 | 101.55 | 0.917 | 0.107 | $2.5 \cdot 10^{-2}$ | -2.7 | -3.9 | 2.7 |
| 17.08 | $7.11 \cdot 10^{4}$ | $2.5 \cdot 10^{14}$ | -4.79 | 88.65 | 1.064 | 0.156 | $2.7 \cdot 10^{-2}$ | -1.4 | 3.9 | 3.8 |
| **17.08** | **$4.94 \cdot 10^{8}$** | **$5.49 \cdot 10^{17}$** | **25.01** | **85.60** | **0.175** | **0.88** | **$7.6 \cdot 10^{-2}$** | **3.5** | **0.5** | **5.9** |
| **17.08** | **$7.12 \cdot 10^{8}$** | **$7.54 \cdot 10^{17}$** | **-10.76** | **96.13** | **1.029** | **3.57** | **$7.2 \cdot 10^{-2}$** | **-2.7** | **1.5** | **2,8** |
| 17.09 | $1.57 \cdot 10^{5}$ | $5.0 \cdot 10^{14}$ | 14.91 | 99.44 | 0.761 | 0.027 | $2.4 \cdot 10^{-2}$ | -2.8 | 7.3 | 7.8 |
| **17.12** | **$1.03 \cdot 10^{6}$** | **$2.6 \cdot 10^{15}$** | **-29.98** | **102.31** | **1.23** | **0.065** | **$1.6 \cdot 10^{-2}$** | **-8.1** | **-1.1** | **6.3** |
| **17.13** | **$3.94 \cdot 10^{9}$** | **$3.4 \cdot 10^{18}$** | **-11.81** | **96.72** | **0.203** | **2.96** | **$7.7 \cdot 10^{-2}$** | **-8.1** | **8.7** | **11.0** |
| **17.15** | **$2.05 \cdot 10^{8}$** | **$2.6 \cdot 10^{17}$** | **-9.77** | **95.88** | **0.551** | **3.28** | **$7.4 \cdot 10^{-2}$** | **-0.7** | **-1.3** | **2.8** |

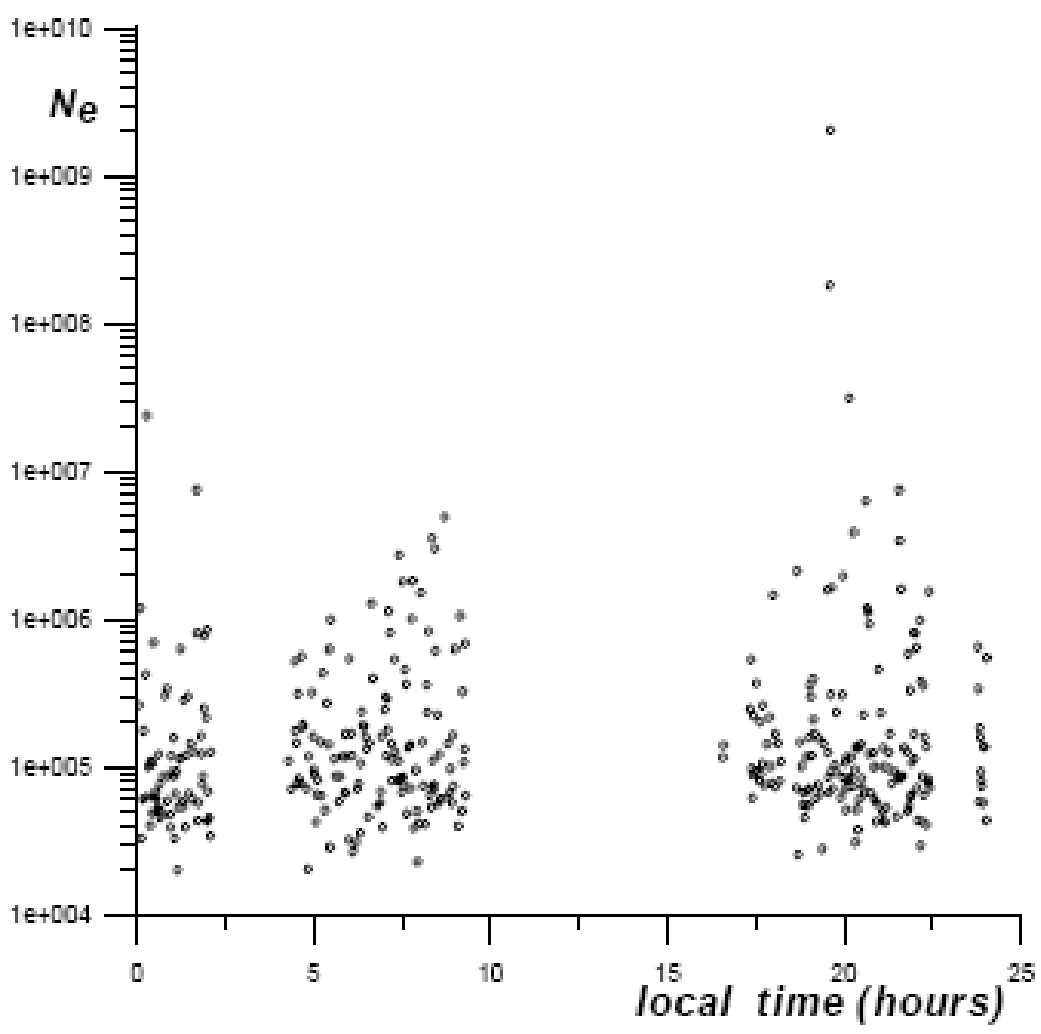

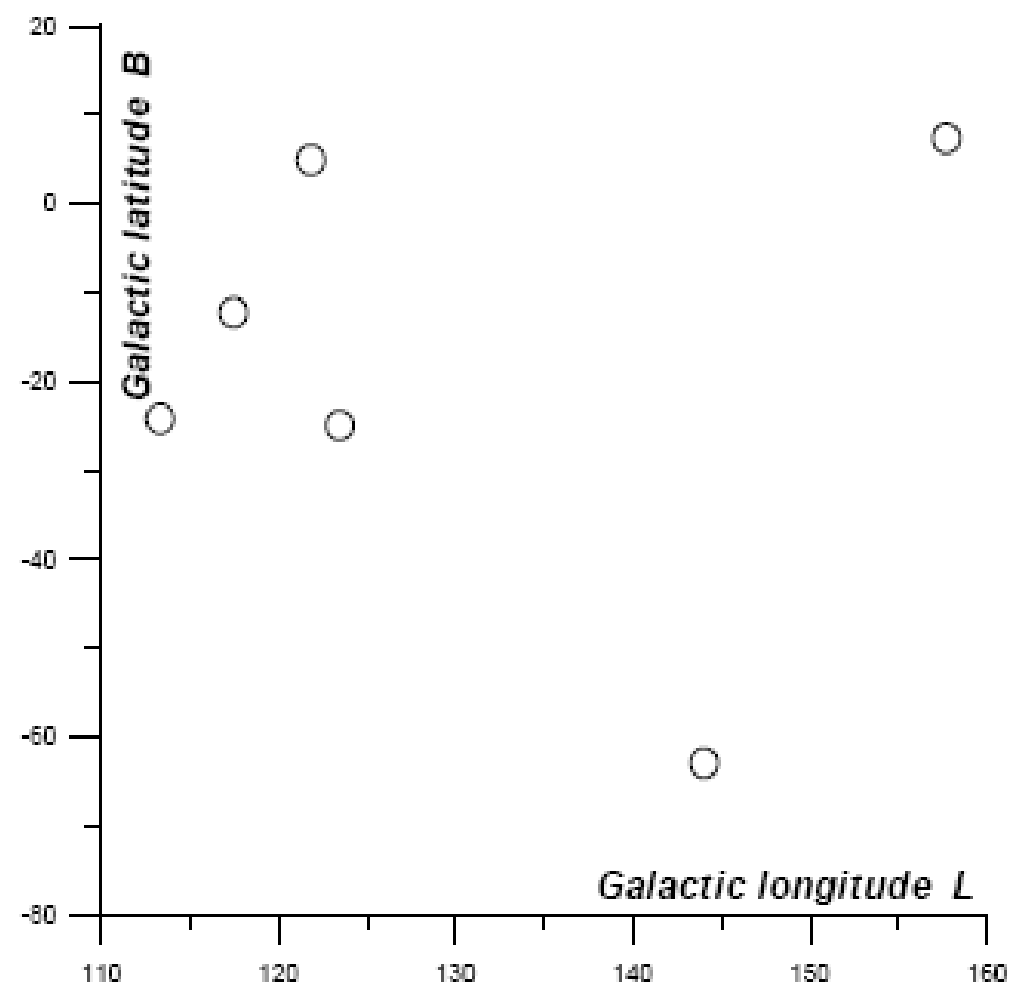

Fig. 5. Series № 5 (Run № 2856), 6 EAS, 26.12.1974, <B> = -13.42°, <L>=117.12°, direction of maximal scattering 110.5° to the Galaxy equator plane. Probability of the occasional observation is < $6 \cdot 10^{-9}$.

Table of data for series № 5 (Run № 2856)

| LOC. TIME | $N_E$ | $E_0$ | B | L | S | EK | FF | X | Y | R |
|---|---|---|---|---|---|---|---|---|---|---|
| **19.53** | **$1.57\cdot 10^{6}$** | **$3.69\cdot 10^{15}$** | **-24.95** | **123.46** | **0.795** | **0.547** | **$2.12\cdot 10^{-2}$** | **-12.4** | **-5.4** | **11.9** |
| 19.55 | $1.23\cdot 10^{5}$. | $4.04\cdot 10^{14}$ | -19.29 | 141.04 | 1.04 | 0.083 | $3.85\cdot 10^{-2}$ | 2.4 | 5.7 | 3.5 |
| 19.56 | $8.5\cdot 10^{5}$ | $2.92\cdot 10^{14}$ | -19.21 | 76.29 | 0.91 | 1.12 | $2.79\cdot 10^{-2}$ | -2.0 | -2.5 | 1.8 |
| 19.58 | $6.65\cdot 10^{5}$ | $1.35\cdot 10^{14}$ | -33.44 | 110.42 | 0.97 | $1.67\cdot 10^{-5}$ | $5.06\cdot 10^{-2}$ | -13.2 | 6.2 | 14.5 |
| 19.59 | 4.11 | $1.55\cdot 10^{14}$ | -3.18 | 122.15 | 0.74 | 0.35 | $3.13\cdot 10^{-2}$ | 3.7 | 2.8 | 3.8 |
| **19.60** | **$5.39\cdot 10^{8}$** | **$5.92\cdot 10^{17}$** | **-24.22** | **113.33** | **0.217** | **0.524** | **$7.55\cdot 10^{-2}$** | **-1.9** | **0.6** | **2.6** |
| 19.60 | $1.04\cdot 10^{5}$ | $3.47\cdot 10^{14}$ | -5.94 | 148.32 | 1.20 | 0.048 | $3.76\cdot 10^{-2}$ | 4.3 | -3.4 | 6.3 |
| **19.62** | **$6.04\cdot 10^{9}$** | **$4.85\cdot 10^{18}$** | **-12.26** | **117.47** | **0.213** | **3.55** | **$7.39\cdot 10^{-2}$** | **-14.1** | **-10.3** | **16.5** |
| 19.62 | $6.55\cdot 10^{4}$ | $2.32\cdot 10^{14}$ | -9.74 | 143.02 | 0.769 | $3.04\cdot 10^{-3}$ | $2.99\cdot 10^{-2}$ | -9.6 | 2.8 | 10.2 |
| 19.62 | $5.05\cdot 10^{4}$ | $1.85\cdot 10^{14}$ | -46.24 | 66.81 | 0.914 | $1.03\cdot 10^{-5}$ | $4.22\cdot 10^{-2}$ | -6.5 | 4.8 | 8.0 |
| 19.63 | $5.38\cdot 10^{5}$ | $1.45\cdot 10^{15}$ | -23.42 | 116.70 | 0.688 | 0.044 | $1.92\cdot 10^{-2}$ | -10.7 | 11.4 | 15.5 |
| 19.63 | $7.43\cdot 10^{4}$ | $2.59\cdot 10^{14}$ | 4.97 | 122.7 | 1.19 | $2.94\cdot 10^{-3}$ | $4.44\cdot 10^{-2}$ | -9.3 | 1.4 | 8.0 |
| 19.65 | $1.1\cdot 10^{5}$ | $3.65\cdot 10^{14}$ | -0.01 | 151.24 | 1.01 | 0.299 | $2.84\cdot 10^{-2}$ | 1.6 | -3.3 | 4.5 |
| 19.65 | $7.7\cdot 10^{4}$ | $2.68\cdot 10^{14}$ | 4.39 | 107.99 | 0.658 | $1.5\cdot 10^{-5}$ | $2.66\cdot 10^{-2}$ | 6.6 | 5.6 | 9.9 |
| **19.66** | **$1.63\cdot 10^{6}$** | **$3.81\cdot 10^{15}$** | **-62.96** | **144.02** | **0.202** | **5.36**$\cdot 10^{-3}$ | **5.96**$\cdot 10^{-2}$ | **-45.8** | **3.6** | **43.7** |
| 19.68 | $7.09\cdot 10^{4}$ | $2.49\cdot 10^{14}$ | -25.43 | 118.33 | 0.497 | 0.021 | $3.64\cdot 10^{-2}$ | -5.5 | -9.8 | 8.5 |
| 19.74 | $1.02\cdot 10^{5}$ | $3.42\cdot 10^{14}$ | -20.96 | 127.44 | 0.843 | $3.23\cdot 10^{-3}$ | $2.84\cdot 10^{-2}$ | 9.19 | -7.2 | 14.8 |
| 19.78 | $2.31\cdot 10^{5}$ | $9.97\cdot 10^{14}$ | -58.39 | 117.39 | 0.761 | 0.025 | $1.89\cdot 10^{-2}$ | -4.6 | 4.2 | 6.7 |
| 19.82 | $9.13\cdot 10^{4}$ | $3.10\cdot 10^{14}$ | -9.77 | 108.68 | 0.99 | 1.93 | $2.38\cdot 10^{-2}$ | -1.55 | -3.2 | 2.0 |
| 19.91 | $6.12\cdot 10^{4}$ | $2.19\cdot 10^{14}$ | -21.67 | 154.65 | 1.165 | 0.035 | $3.80\cdot 10^{-2}$ | 5.51 | 1.4 | 5.8 |
| 19.96 | $3.04\cdot 10^{5}$ | $8.84\cdot 10^{14}$ | -24.72 | 146.23 | 0.813 | 4.81 | $2.07\cdot 10^{-2}$ | -1.33 | 6.5 | 1.6 |
| **19.98** | **$1.94\cdot 10^{6}$** | **$4.43\cdot 10^{15}$** | **7.37** | **157.69** | **0.662** | **0.185** | **$1.52\cdot 10^{-2}$** | **3.8** | **22.2** | **14.6** |
| 19.99 | $8.42\cdot 10^{4}$ | $2.89\cdot 10^{14}$ | 13.66 | 136.92 | 1.023 | 0.062 | $3.01\cdot 10^{-2}$ | -7.8 | 1.6 | 4.7 |
| 20.01 | $6.84\cdot 10^{4}$ | $2.41\cdot 10^{14}$ | -32.65 | 136.60 | 1.169 | 0.076 | $3.39\cdot 10^{-2}$ | -1.6 | 1.1 | 1.6 |
| 20.05 | $1.46\cdot 10^{5}$ | $4.66\cdot 10^{14}$ | -41.61 | 118.12 | 1.82 | 0.282 | $2.99\cdot 10^{-2}$ | 4.8 | -6.7 | 7.5 |
| 20.05 | $6.65\cdot 10^{4}$ | $2.35\cdot 10^{14}$ | 0.77 | 134.65 | 0.955 | 0.156 | $2.70\cdot 10^{-2}$ | 2.7 | 4.4 | 3.9 |
| 20.06 | $5.07\cdot 10^{4}$ | $1.86\cdot 10^{14}$ | -37.07 | 136.51 | 0.896 | 0.284 | $3.21\cdot 10^{-2}$ | -2.0 | 1.5 | 1.8 |
| 20.08 | $1.15\cdot 10^{4}$ | $3.81\cdot 10^{14}$ | -35.59 | 119.79 | 0.884 | $8.31\cdot 10^{-3}$ | $2.18\cdot 10^{-2}$ | 6.2 | -7.7 | 8.6 |
| 20.09 | $5.37\cdot 10^{5}$ | $1.95\cdot 10^{14}$ | -36.25 | 157.03 | 1.147 | 0.321 | $5.07\cdot 10^{-2}$ | 5.2 | 3.3 | 1.7 |
| 20.09 | $9.44\cdot 10^{4}$ | $3.19\cdot 10^{14}$ | -14.61 | 118.45 | 0.674 | 0.029 | $2.01\cdot 10^{-2}$ | -8.5 | 3.4 | 10.0 |
| 20.11 | $8.12\cdot 10^{4}$ | $2.80\cdot 10^{14}$ | 8.71 | 118.57 | 0.643 | $4.82\cdot 10^{-3}$ | $2.61\cdot 10^{-2}$ | 4.3 | 7.1 | 7.3 |
| **20.17** | **$3.1\cdot 10^{7}$** | **$4.94\cdot 10^{16}$** | **4.91** | **121.84** | **0.207** | **0.256** | **$1.66\cdot 10^{-2}$** | **29.9** | **-6.8** | **29.6** |

As of to day, no other series, satisfying the requirements stated to the quality and reliability of the events as well as to all above given conditions applied to the duration and the energy of events, were found in the material at our disposal. Nevertheless the additional treatment of these events and the search for the series will be continued.

Very urgent is to be aware that the registered phenomenon is not an imitation, namely that the EAS in series are true extensive air showers. One can bring three arguments in favor of the registered EAS veritablity.

1. There are in total 110 EAS in the time frames of all registered series (including the EAS of comparatively small energies). 100 of them are accompanied by the nonzero energy release in the ionization calorimeter. In addition, in 77 events there are distinct penetrating tracks in the calorimeter, in many cases multiple ones (pairs, threes) are registered, with the accompanying cascades. As an example, the events in the calorimeter for one of the series - № 1 were shown above. There are several large nuclear cascades as well, two of them are shown in Fig. 6.

12.62 (Run № 2746)

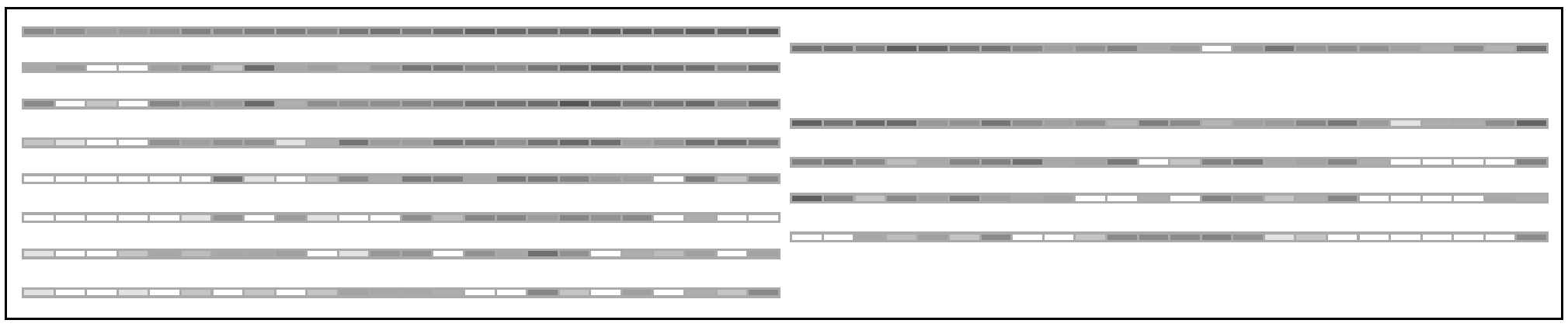

19.62 (Run № 2856)

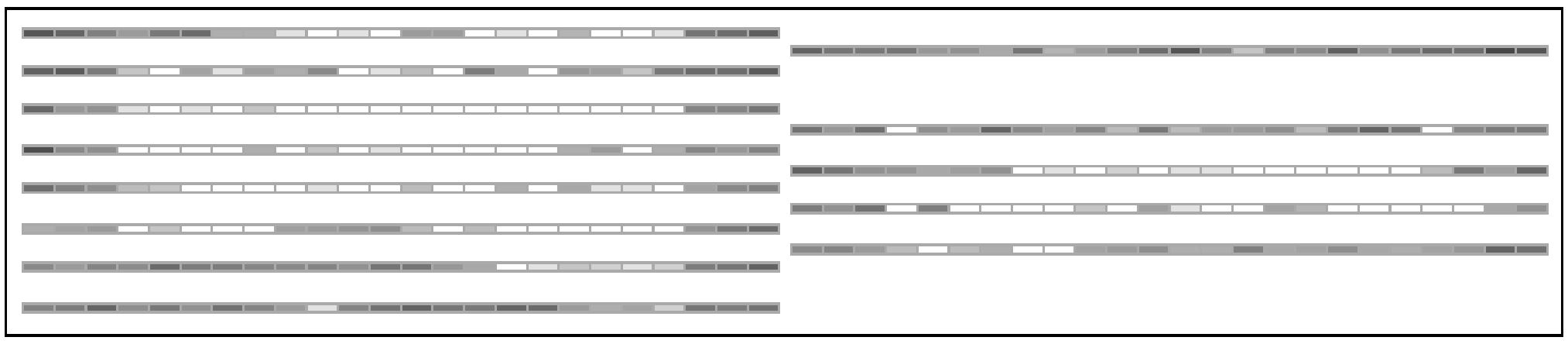

Fig.6. Examples of large nuclear cascades in calorimeter.

Only 10 out of 110 events have no accompanying hadrons in the calorimeter (EK <~ $2.0 \cdot 10^{-5}$. The axes of 6 of these events are at large distances from the calorimeter (>~ 10 m). It is not excluded however, that at least some part of hadron less events is initiated by primary gamma-quanta (criteria of the possible gamma-EAS selection are discussed in [3]).

2. Comparison of fit level FF to NKG distribution for the observed events, with that for the events outside the series. Before the observation of the series, by the work on fitting of EAS to NKG function, it was found that the distribution of FF is located in the borders 0.02 <= FF <= 0.12 and has a two-hump form - with maximums at the fit levels FF=0.03 and FF=0.07, with a distinct trough between them and with the maximal value of FF = 0.12. This distribution for $N_e > 10^6$ is shown in Fig. 7.

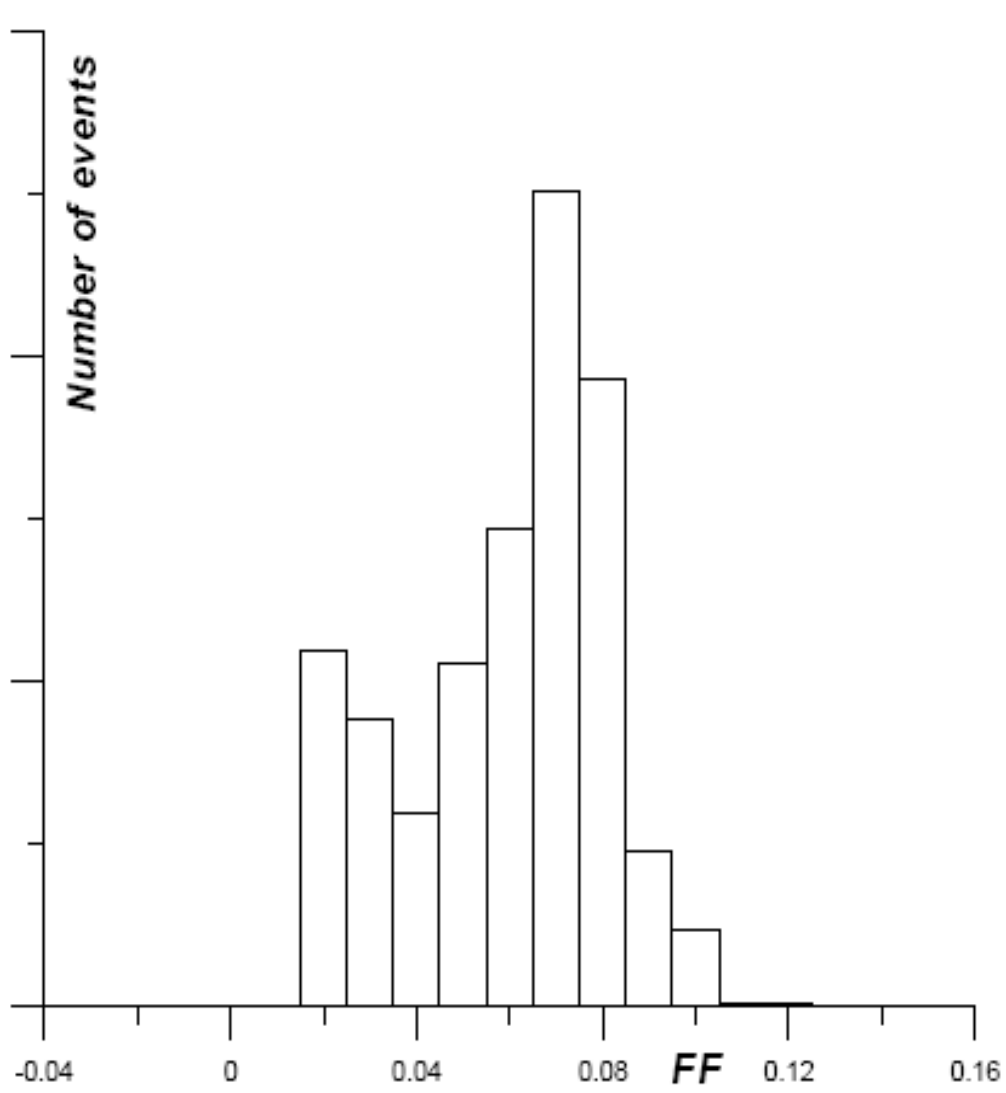

Рис. 7. Distribution of squared deviation FF of NKG function from experimental spatial distribution in the individual EAS outside the series for $N_e > 10^6$.

The events of the series in whole diapason $N_e >\sim 10^6 - 10^9$ form a completely analogous distribution of FF with domination in the region of the second maximum, i.e. around of FF=0.07, like the events of the analogous energies outside the series, not crossing the limits, characteristic for the “ordinary” EAS – see the Tables of data for series. What defines the two-hump character of this distribution is not clear yet. It seems to be natural to explain the presence of the second hump by the contribution of more energetic EAS, which may have several secondary axes. These EAS naturally do not fit so well. The main thing here is that the fit level of the EAS of series does not exceed that of the ”ordinary” EAS.

3. In ~70% of cases, the quality of the events in the calorimeter makes it possible to estimate the spatial angle of the EAS registration, independently from the chronotron part of the installation. For about 50 events, it is possible to estimate the spatial angle by means of both methods simultaneously – through the chronotron and through the hadron component data in the calorimeter. We stress, that these two approaches are completely independent, both in the sense of device and of the measurement method. In Fig.8 the distribution of the differences of the zenith angles obtained via these 2 methods in parallel is shown. For more reliability, the distribution is built for the angular diapason $20^0 - 60^0$, excepting the interval $0^0 - 20^0$. In the majority of cases, the closeness of the results obtained by means of these two methods is obvious.

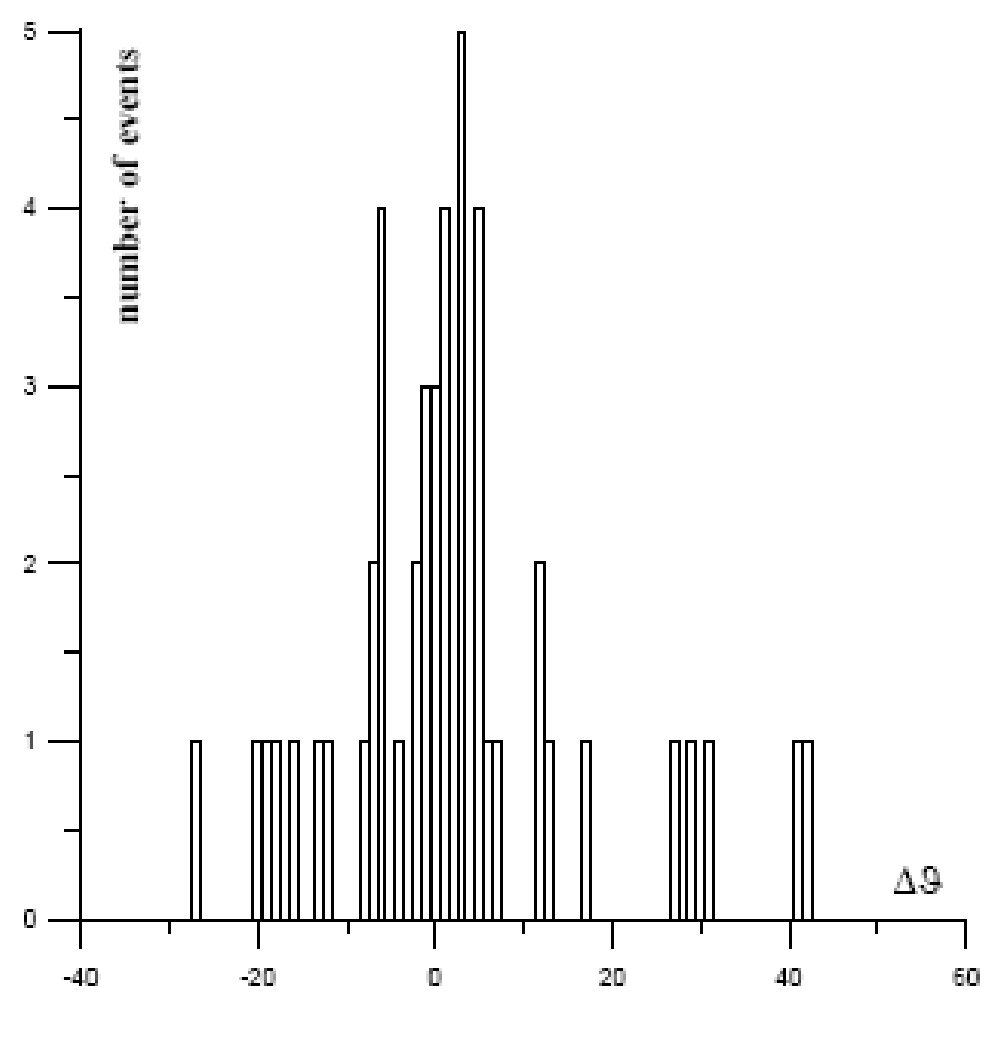

Fig .8. The distribution of the differences of the zenith angles, measured via the chronotron and via the hadron component data in the calorimeter in parallel, for the angular diapason $20^0 - 60^0$

A similar situation is observed for the azimuth angles as well (Fig.9). It is obvious, that the precision of the azimuth angle estimation must be less compared with that for the zenith angle. For example, for the very small zenith angles (less then the precision of their estimation) the possible error in azimuth can reach $180^o$.

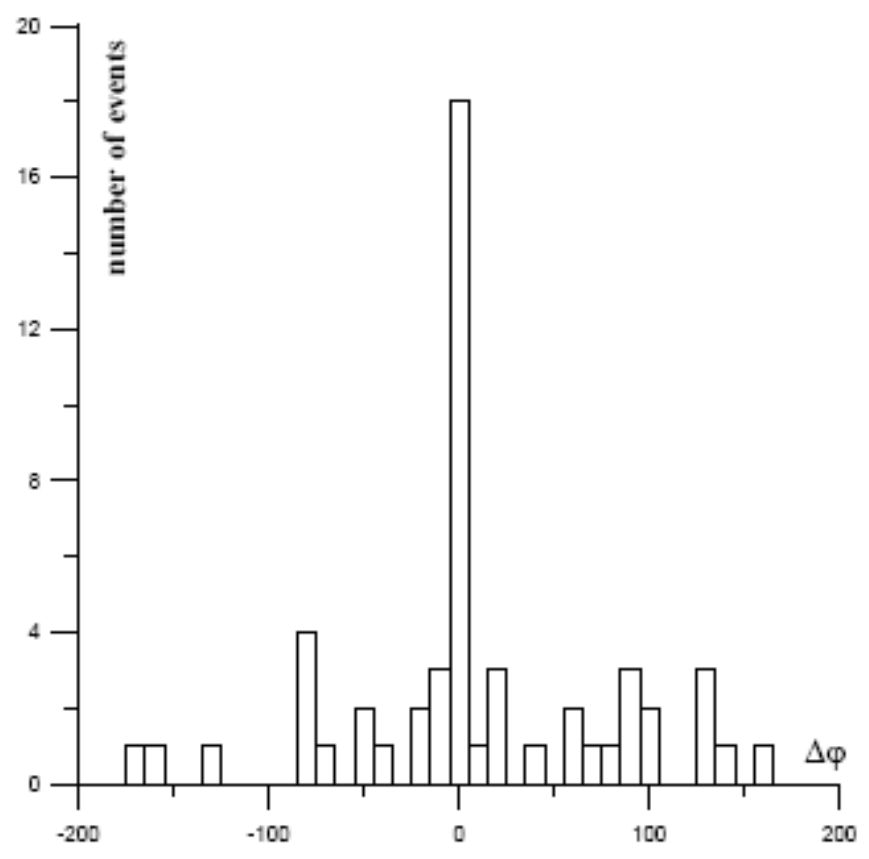

Fig. 9. The distribution of the differences of the azimuth angles measured via the chronotron and via the hadron component data in the calorimeter in parallel. (The events are the same, as in Fig.8).

Note, that in the basic treatment of the data: the angles measured both by means of the calorimeter and of the chronotron, are taken into account, and then the event is fitted to the NKG function according to the indications of all transducers of the installation.

So there are all reasons to be certain, that the observed events are of good quality by all parameters; the behavior of their hadron and electron components corresponds each to other and to commonly accepted characteristics: and so these events evidently are true Extensive Air Showers.

It is not excluded, that part of the EAS of relatively small energies, located in the same time interval, is genetically connected with “their” series but is lost against the background of the rest of events coming with high intensity. In favor of this speaks the comparison of the galactic coordinates of events registered inside the borders of the time interval of the series – see Tables. In some cases these EAS may be traced visually, for example in the series № 1 (run № 2342) and № 3 (run № 2751), though it is rather subjective. It will be the object of further consideration.

Note, that the series of Gamma - EAS in the region of relatively low energies $N_e \sim 10^{15}$ eV were observed earlier. As an example one can quote the large event, consisting of 32 gamma-EAS, observed in the University of Manitoba [4]. The event was registered at 20.01 1981. Unfortunately there are no experimental data from our installation for this data at our disposal.

The characteristic peculiarity of all observed series is the asymmetry of the galactic coordinate distribution. To estimate the direction of this asymmetry, for each series the galactic coordinates <L> and <B> of the centre of weight were estimated – using the EAS primary energy $E_0$ for the weight of the event. Next, the peculiar axis of symmetry is searched for, which crosses this centre of weight and with respect to which the mean square deviation over the whole series takes the minimal value, i.e. the direction of this axis reflects the direction of maximal scattering of the events in the series under consideration. It is easy to see that two series - № 4 (run № 2855) and № 5 (run № 2856 ) - are coming from directions close to the galactic equator plane – galactic latitudes <B> of the centers of weight $-6.05^o$ and $-13.42^o$ correspondingly. For these two series the direction of maximal scattering is close to normal to the galactic equator plane - 94.5$^o$ and 110.5$^o$ correspondingly. Regarding the magnetic fields as the main cause of the scattering, one can see that this observation is in accordance with the predominant direction of the magnetic field near the galactic equator plane along this plane (i.e. normal to the direction of the maximal scattering). In two series - № 2 (run № 2746) and № 3 (run № 2751) centers of weight are located far enough from the galactic equatorial plane (galactic latitudes 58.57$^o$ and 60.47$^o$ correspondingly), and the direction of maximal scattering is close to parallel with the galactic equator plane (10.5$^o$ and 170.5$^o$ degrees correspondingly). This is in favor of the predominant direction of the magnetic field being close to normal to the galactic equator plane for large galactic latitudes. For series № 1 (run № 2342 ) the center of weight is again close to the galactic equator plane ($\langle B \rangle = -2.14^o$ ), however the direction of maximal scattering has here the intermediate value 34.5$^0$ , which possibly means, that the average magnetic field along the direction of the mean trajectory of this series has the essential slope to the galactic equator plane.

## CONCLUSIONS

1. The bursts of EAS - the series of the high and super high energy EAS following each other with the mean time interval between them 1 – 5 minutes were observed and investigated. The duration of the each separate series is <~ 30 minutes. It is essential, that the frequency of the appearing of such series is much higher, than the probability of the occasional fluctuations in the succession of independent EAS. This probably is in favor of the commonality of their origin.
2. All the basic characteristics of EAS forming these series are obtained.
3. Apparently, the majority of events in these series are initiated by the charged particles, and not by the gamma quanta. This is indicated by asymmetry of the galactic coordinate distribution of the axes directions of EAS, composing these series.
4. The approach described may serve as an additional method for the search of cosmic radiation sources, and for the independent evaluation of magnetic fields distribution in the Galaxy. In particular, in presence of identified source of the series, with known distance to it, the evaluation of the intensity of the magnetic field along this direction becomes possible.
5. We do not discuss the possible nature of the phenomenon described. Just one remark. If one supposes that EAS of the series were emitted isotropic from the source, the source must be taken as being located close to the Earth (at the distance of the order of several thousands of light years). Otherwise unlikely huge energies must have been emitted. It is possible to suppose the jet-like character of the emission.
6. The possibility seems to be interesting, that at least the part of the described series could have been observed on other installations at the quoted dates and time. In this case it would be useful to carry on further investigations with joint efforts. In particular it could be possible to use the analogous data from other installations to search for synchronous events in our material.

ACNOWLEDGEMENTS

The authors are sincerely grateful: to A.D.Erlikin, for fruitful discussions and advises, to V.P.Pavluchenko for useful discussions, concerning the data bank and installation, to O.V.Kancheli for his constant attention to our investigations, well-meant critical remarks and advices, to J.M.Henderson for useful and interesting discussions and his assistance in preparation of this paper. We are grateful to the astronomer from the Abastumani Observatory E.Janiashvili for consultations regarding the astronomical calculations. We are grateful to firm Manta Systems ltd. (Tbilisi) for the use of their calculating powers.